\documentclass[fleqn,10pt]{wlscirep}
\usepackage[utf8]{inputenc}
\usepackage[T1]{fontenc}
\usepackage{subcaption}
\usepackage{float}
\usepackage{multicol}
\usepackage{multirow}
\usepackage{pdflscape}

\title{Social Cyber Geographical Worldwide Inventory of Bots}

\author[1,*]{Lynnette Hui Xian Ng}
\author[1]{Kathleen M. Carley}
\affil[1]{Center for Informed Democracy \& Social - cybersecurity (IDeaS), Societal and Software Systems
Carnegie Mellon University, Pittsburgh, PA 15213}

\affil[*]{lynnetteng@cmu.edu}


\begin{abstract}
Social Cyber Geography is the space in the digital cyber realm that is produced through social relations. Communication in the social media ecosystem happens not only because of human interactions, but is also fueled by algorithmically controlled bot agents. Most studies have not looked at the social cyber geography of bots because they focus on bot activity within a single country. Since creating a bot uses universal programming technology, bots, how prevalent are these bots throughout the world?
To quantify bot activity worldwide, we perform a multilingual and geospatial analysis on a large dataset of social data collected from X during the Coronavirus pandemic in 2021. This pandemic affected most of the world, and thus is a common topic of discussion. Our dataset consists of $\sim$100 mil posts generated by $\sim$31mil users. Most bot studies focus only on English-speaking countries, because most bot detection algorithms are built for the English language. However, only 47\% of the bots write in the English language. To accommodate multiple languages in our bot detection algorithm, we built Multilingual BotBuster, a multi-language bot detection algorithm to identify the bots in this diverse dataset. We also create a Geographical Location Identifier to swiftly identify the countries a user affiliates with in his description. Our results show that bots can appear to move from one country to another, but the language they write in remains relatively constant. Bots distribute narratives on distinct topics related to their self-declared country affiliation. Finally, despite the diverse distribution of bot locations around the world, the proportion of bots per country is about the same ($\sim$20\%). Our work stresses the importance of a united analysis of the cyber and physical realms, where we combine both spheres to inventorize the language and location of social media bots and understand communication strategies.
\end{abstract}
\begin{document}

\flushbottom
\maketitle

\section*{Introduction}
Social interactions between agents in the cyberspace are reflective of offline communities and regional boundaries. ``Social Cyber Geography" is the physical geography of the social communities that arise from the cyber realm. For example, visualizing the telecommunication map reveals the physical boundaries between Wales, England, and Scotland \cite{ratti2010redrawing}. Many tools allow us to construct social cyber geographies. Location-based social networks provide a plentiful source of geolocated connections between individuals \cite{andris2016integrating}. Analyzing social cyber geographies help us to have more precise response to events. For example, creating a social cyber geography of communications during the coronavirus pandemic revealed the level to which the countries were affected and indicated where resources were needed \cite{qazi2020geocov19,edry2021real}. More so, in the area of social cybersecurity, which focuses on understanding how behavior is influenced by relationships and communities and the analysis of information maneuver campaigns \cite{carley2020social}, creating a social cyber geography of these campaigns reveals patterns of global alliances and threats. For example, the social cyber geography of bots and their stances during the 2022 Russian-Ukraine war revealed the distribution of political stances on X towards each country \cite{shen2023examining}.
Here, we bridge the cyber realm and the physical world, by creating a social cyber geography of social bot agents.

Our focus on bot agents builds on the spate of past work analyzing these automated agents on social media, many of which declare that these agents have malicious agendas such as spreading disinformation. Bot agents have been key agents in the dissemination of coronavirus misinformation \cite{ng2024exploring} and negative vaccine stance \cite{unlu2024unveiling}, influence elections around the world \cite{ferrara2020bots,ng2024tiny}, and take part in influencing people's stances \cite{duan2022algorithmic,ng2022pro}. Bot agents are also disproportionately distributed on the geographic map, indicating their intention towards specific population groups \cite{tan2023botpercent}. 
Therefore, it is important to analyze the distribution of declared affiliation and language used of bot accounts to be able to better understand their objectives in cyberspace, such as who they are targeting with their narratives. 

Social Cyber Geography provides a worldwide inventory of social bots. This inventory allows analysis of communities of bots. A community of bots is a set of bots who claim to be from the same country in their metadata (i.e., user description) and are tweeting in the same language. Our social cyber geography allows analysis of the distribution of the desired affiliation of bot agents, the language use across the globe by bots, and the target audience of the disseminated narratives. The social cyber geography view reflects the communicative strategies of bots and insights towards the use of automation for information dissemination on social media.

This social cyber geography is created from a large dataset consisting of almost 31 million users collected from X (formerly known as Twitter). These users tweeted posts related to the 2021 coronavirus pandemic. Methodologically, we create a multilingual bot detection algorithm and a gazetter-based geolocation identifier that leads to the construction of this social cyber geography. The multilingual bot detection algorithm is necessary to identify bot agents across diverse languages, since current bot detection algorithms mostly support the English language, and thus miss linguistic cues in other languages that characterize a bot \cite{albadi2019hateful}. The gazetter-based geolocation identifier relies on NLP techniques rather than API calls to geo-databases as a solution to identifying locations of users at scale.

By being able to geographically analyze bot activity, we now understand the extent that bot activity is a worldwide phenomenon, and the regions where bots are used to support native speakers of the country as opposed to communities of immigrant diasporas.

\section*{Results}
By segregating the social bot discourse by their geography, we find that bots are present in almost every part of the world that discuss the coronavirus pandemic. In fact, \autoref{fig:bot_perc_country} shows that regardless of the country, the average proportion of bots is approximately 20\%. 

An over-time investigation (see Supplementary Material \autoref{fig:bot_perc_country_monthly}) shows that geographical affiliations change over the time of the coronavirus discourse. Countries that do not have permanent human habitation like Antarctica and countries where X is restricted like China are also used as affiliations, where agents claim they originate from those countries. Notably, China was heavily affiliated at the beginning and middle of the pandemic. The start of the pandemic reveals many accounts affiliating with China but are likely to be from America (e.g., @DemSocialist). Manual inspection reveals that these agents are news bots providing updates on case counts and malicious bots promoting anti-Chinese hate. The FDA gave full approval for the Pfizer-BioNTech vaccine in the middle of the pandemic (Aug 2021), sparking many accounts affiliating with China to promote the Sinovac vaccine over Pfizer.
The affiliation of social bots by geography reflects the employment of digital diplomacy, where narratives are portrayed to be originating from a country, and agents leverage the anonymity of social media to hide their true origin. Positive digital diplomacy stems from the use of narratives to sway public opinion to the country's advantage (i.e., promote home vaccine), while negative digital diplomacy paints other countries in bad light \cite{gilboa2013public}. 

We characterize the nature of discussions through the language that bots use. Asian and European languages have the highest percentage of bots. Among Asian and European languages, those with the highest percentage of bots are: Thai (59.5\%), Japanese (28.3\%), Tamil (22.6\%), Portugese (18.7\%), Greek (28.9\%), Lativian (37.5\%). In comparison, the bot percentage of the English language is 14.7\%. Tweets in English are general discourse pertaining to the coronavirus vaccine, lockdown information, news on the pandemic, and key events or symptoms to watch out for. Tweets in European languages promote a sense of medical hesitancy (translated from Spanish, ``[...]Starting in June, cases of myocarditis appeared among young people after receiving the vaccine, mainly among men[..]"). Tweets in Asian languages educate people on the side effects of the coronavirus vaccine (translated from Japanese with Google Translate: ``No one is saying that vaccines can suppress symptoms. It has a certain degree of effectiveness in preventing infection, onset, and aggravation. This does not mean that after-effects do not occur after a breakthrough infection. And the only indicator of “severe” COVID-19 is pneumonia symptoms"). The language in which the bot writes reflects its target audience -- users who can read the language. Bots write different narratives in different languages, revealing how they are used globally for strategic communications.

As a deeper dive into language distribution, we compare the bot distribution against the dominant language of a country. Despite the changing distribution of the bot affiliation, the distribution of language used is rather consistent across the year-long data. On average, at least 80\% of the bots affiliated with a particular country use the dominant language of the country. This indicates that bots target language diasporas across countries. There are nine countries where less than 80\% of the bots affiliated with the country use the country's dominant language. When the dominant language is not used, the bots use Asian and European languages (\autoref{tab:countries_languages_less_main}), which agrees with the observation that Asian and European languages have the highest percentage of bots.

Bot narratives are pertinent regional narratives. \autoref{tab:dominant_hashtags_main} shows the top hashtags coming from posts geolocated to different region of the world (United States, Canada, Russia, Africa, India, Indonesia, China, Asia). Tweets geolocated to the United States focus on the vaccine and masks. Tweets in Canada and India were more politically focused, the former on the 2022 general elections, the latter on their Prime Minister. Tweets geolocated to Russia promote Russia's SputnikV vaccine, indicated by top hashtags like ``AWANIpagi" (early morning) and ``NormaBaharu" (new normal), which are hashtags written in Indonesian and Malay. These hashtags are repeated in the Indonesian set of hashtags, which focus on the morning news and the vaccination police. Interestingly, the hashtags geolocated to China ``UnrestrictedBioweapon" and ``ContemporaryGeneticWeapon", indicate deliberate and automated blame by these bot agents on the Chinese for the coronavirus outbreak.
Tweets geolocated to the Africa region prey on human fear by comparing the coronavirus with previous pandemics of Ebola and HIV, and also call for VaccineEquity for the undeveloped region.

Comparing bot geographical activity against country indicators reveals no significant correlation between bot activity and economic activity or population across geographical space. This indicates that people from all countries are equally likely to create bots and all countries are equally likely to be used as affiliations for bots. One possible reason for that is that the country that constructs or pays for the bot is not the one the bot is affiliated to. During the coronavirus pandemic, governments use bots to send messages, news agencies to broadcast regional case counts, and for marketing purposes to promote health-related products or spread (dis)information \cite{ng2024cyborgs,chang2021social,hajli2022social}. 
Finally, we note some key areas where there are very few bot agents: Mexico, Central and northern Africa, and the Myammar regions. One explanation is that the combination of a later onset of the virus in Mexico and the African regions compared to the rest of the world and their younger population means that the regions were relatively unaffected by the virus, and thus have less of a discussion of the virus on social media \cite{dzinamarira2020covid}. 
The second reason is that, as reported by Statistica, there is a very low use of X. The average number of users in these regions is only 0.6 million users. Therefore, it is reasonable that bot agents do not affiliate nor target these regions, since the audience reach is rather small. 
In the Myammar region, there is a third reason, which is political turbulence, where the military overthrew the democratically elected government. During this period, X was temporarily blocked, which caused discourse to dwindle further and users to move away from the platform.
These observations mirror previous work which mapped organically geolocated tweets during the coronavirus pandemic, and showed a very low number of tweets (0-1000) in these geographical regions \cite{qazi2020geocov19} (see Supplementary Material \autoref{tab:comparison_with_geocov19} for a detailed comparison).

\section*{Discussion}
Through the synergy of a multilingual bot detection model (\emph{``Multilingual BotBuster"}) and a geographic location identifier (\emph{``Geolocation Identifier"}), we built a social cyber geographical worldwide inventory of bots. This inventory reveals that social media bot agents are spatially distributed across all countries around the globe, with a higher concentration in the more politically charged countries (e.g., China, United States). This spatial affiliation of bot agents changes over time, indicating the bots' shifting appearances of origin. Despite shifts in allegiances, the bot percentage in each country's dominant language stays roughly the same, indicating that there is a set of readers that the bots consistently target. Overall, some languages are more likely to be used by bot accounts, especially Asian and European languages; and narratives written in each language are tailored to the speakers of those languages.

We extend previously developed bot detection algorithms \cite{ng2023botbuster,yang2022botometer} towards a multilingual algorithm in order to more accurately detect bot accounts. Our bot detection algorithm allows detection of bot users who write in languages other than English. We create a gazetteer-based Geolocation Identifier to infer country affiliation through the bot's user description. Our bot inventory reveals that bots are present in all languages and locations, and mostly write in the dominant language of the country they affiliate with. This multilingual bot detection algorithm is important because it reveals bots that were previously undetected by English-based bot detectors.  Our results are consistent with previous work on the geographical distribution of tweets \cite{tan2023botpercent,qazi2020geocov19}. We extend those works to show that the percentage of bots is comparable across regions, and the exact number is proportionate to the number of tweets. Further, our time-series analysis shows that on the global scale, bot affiliation with countries and language is time-invariant.

Social media recommendation algorithms show users' posts that are in the language they declare and around the region they are geolocated to. Our bot inventory goes beyond a recommendation system. The inventory provides a cross-country survey view that bot communications are strategic. For the coronavirus pandemic, bots target regions that are heavily affected by the virus, and use the region's dominant language to maximize their reach. In fact, this social cyber geography shows how bots are actively using strategic communications: bots that claim to be from China disseminate narratives blaming China for the coronavirus. At the same time, the Chinese state-sponsored media was actively denying that the virus came from China, and that it was actually coming from the US \cite{nadesan2022crises}. This is a form of social cyber attack that aims to influence the Chinese public about the source of the virus. Bot influence is therefore an international phenomenon not confined to a certain country. An international consortium might be useful to regulate bot activity worldwide, set guidelines for when bot activity is excessive, and become influence operations such that the bot agents need to be regulated. 

Several limitations nuance this work. Crucially, the country annotation relies on the comparison of a gazetteer with the agent's declared location. The gazetteer should be constantly updated with the latest social media location lingo, and location identification might be extended to analysis of the text, rather than only the description. Further, the geographical affiliation derived is where the bots claim they are from, not where they are actually from. To determine the location where the bots originate from requires techniques like IP tracing, which is beyond the scope of the paper, and future work should explore if the affiliation were native or due to external influences (e.g., use of VPN). However, the declared affiliation shows the location the bot desires to appear coming from, which is the information that human users will process when they scroll their social media feed.

The Worldwide Inventory present in this paper has broad coverage of geographical regions and languages, which we believe can foster downstream multidimensional research. Future work involves characterizing these bots into different types of bots and analyzing the geographical distribution between bot types. Such an analysis provides insight towards the affiliations and languages that each type of bot tends to adopt, and thus provides evidence for inference of purposes of each bot community. Importantly, bot agents are AI agents, and are built for the wide creation and dissemination of misinformation during the coronavirus pandemic \cite{himelein2021bots}, a topic which the WEF (World Economic Forum) named as the top threat in 2024. Our inventory facilitates the analysis of bot-distributed misinformation by region. 

\section*{Conclusion}
Automated information dissemination by bots through social media crosses physical geographical boundaries. Social Cyber Geography of bot activity is needed to unmask the bots' social targets and bot messaging techniques. The combination of a multilingual bot detection and geolocation of individual messages provides information on how automated bot agents target and share information within the physical geographic landscape. Our worldwide bot inventory reveals the social cyber geography of bots, pointing out social affiliations and targeted communities, leading investigations and regulations towards objectives of bot agents through the narratives they put forth.

\begin{figure*}%[tbhp]
\centering
\includegraphics[width=0.8\linewidth]{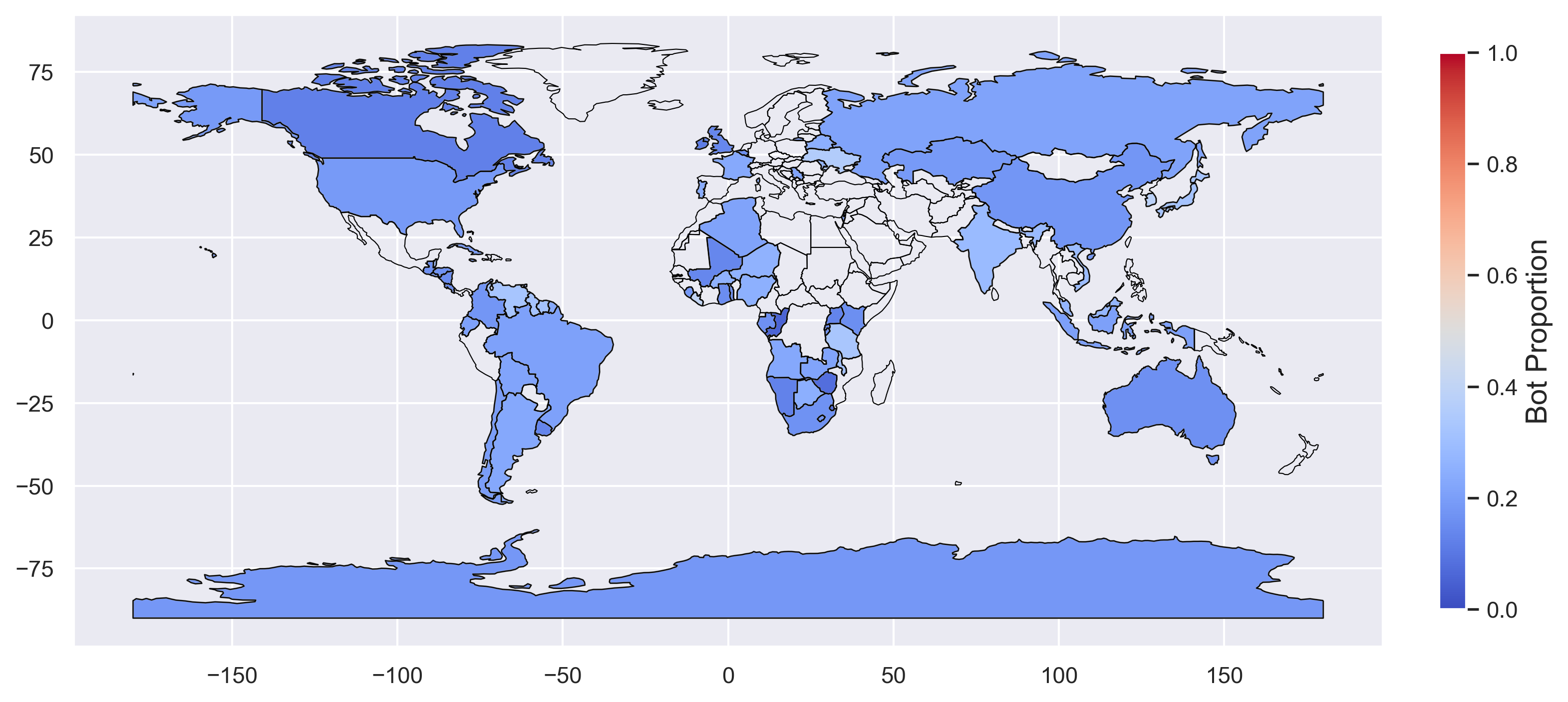}
\caption{Geographic Heat Map of the average (median) percentage of bots affiliated with each country, across the entire data. White areas indicates that there are no bots present in the data we collected.}
\label{fig:bot_perc_country}
\end{figure*}

\begin{figure*}%[tbhp]
\centering
\includegraphics[width=0.8\linewidth]{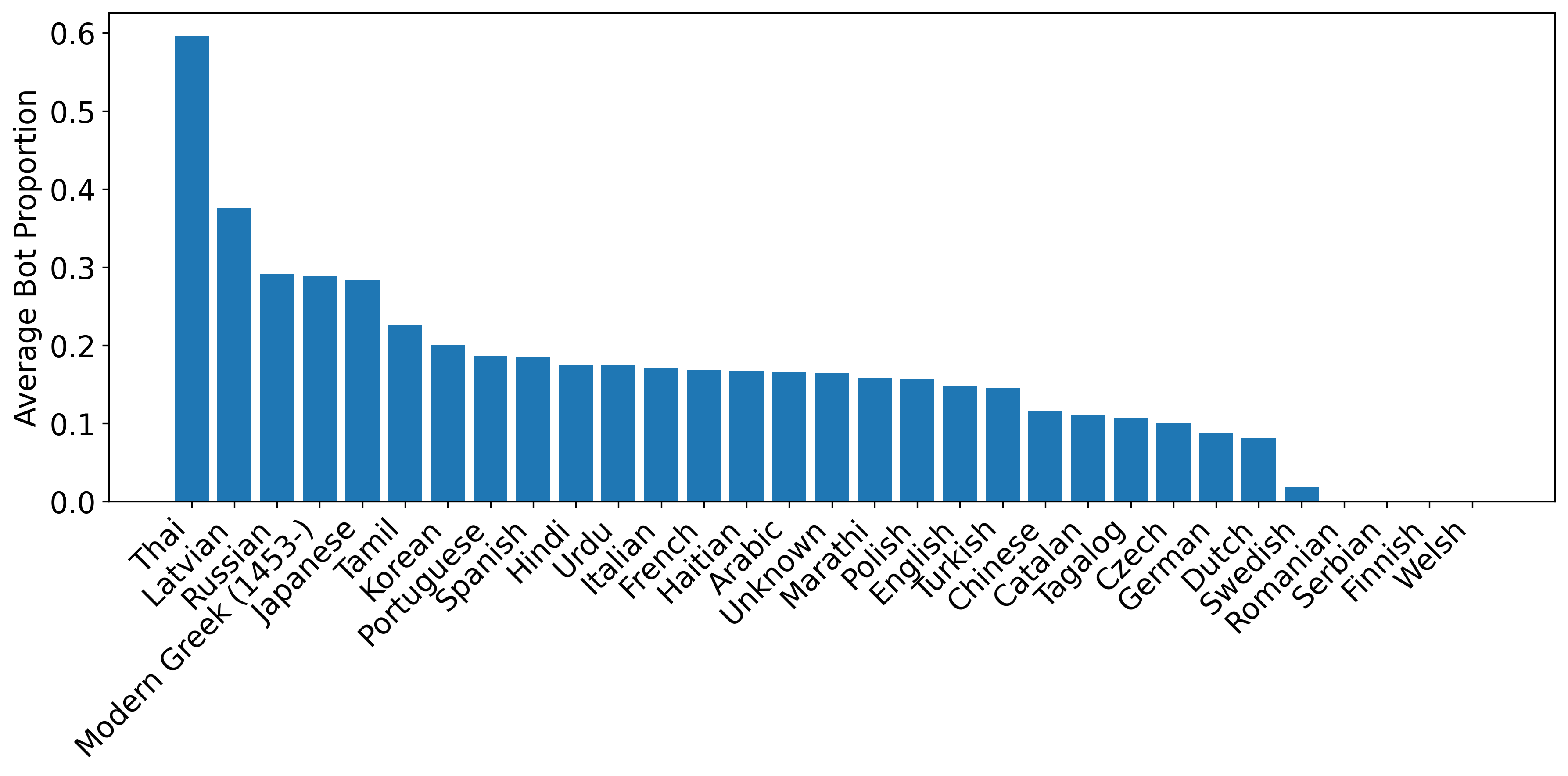}
\caption{Languages that have the highest proportion of bot users.}
\label{fig:bot_perc_language}
\end{figure*}

\begin{figure*}%[tbhp]
\centering
\includegraphics[width=0.8\linewidth]{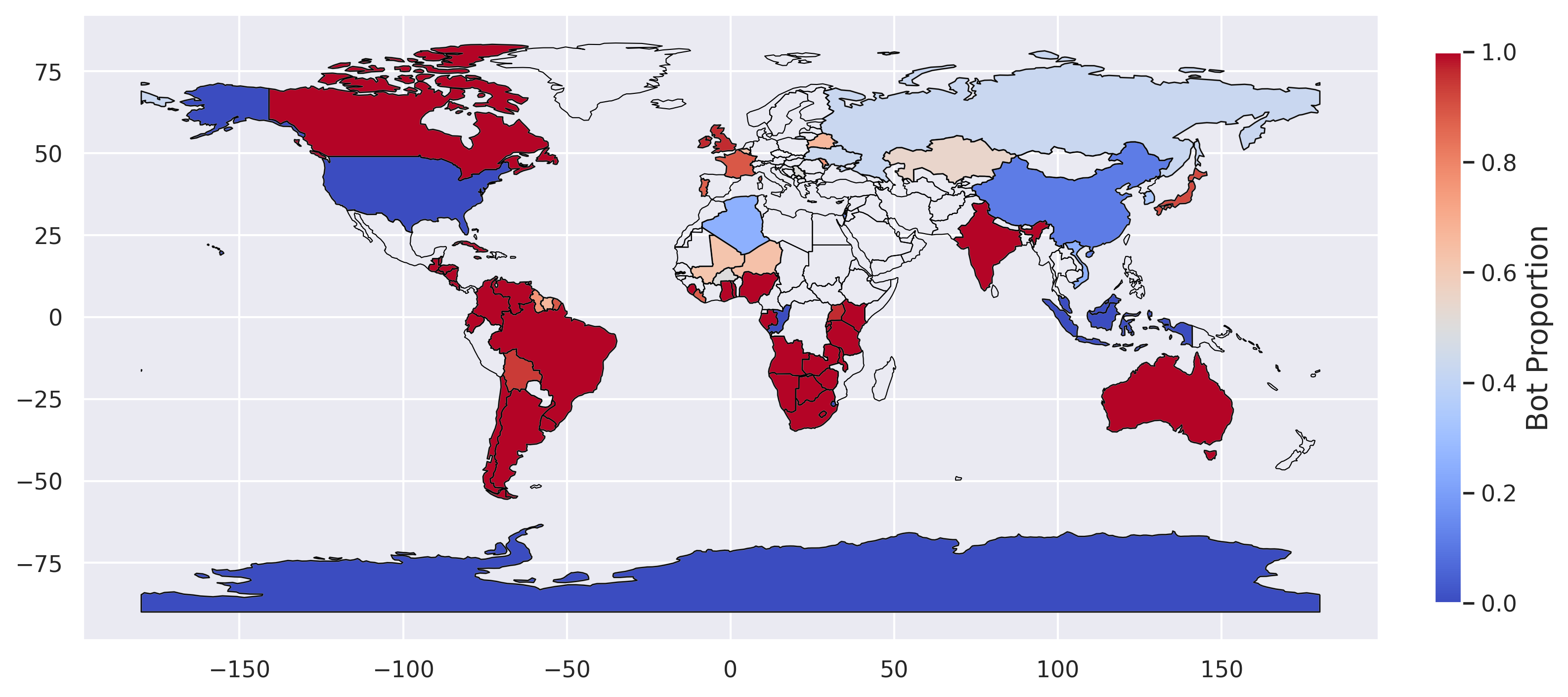}
\caption{Geographic Heat Map of the average (mean) percentage of bots affiliated with a country that authored posts in the country's dominant language, across the entire data. White areas means that there are no bots present in the data we collected.}
\label{fig:bot_perc_lang_dominant}
\end{figure*}

\begin{table}[h]
    \centering
    \begin{tabular}{p{2cm}p{5cm}}
    \toprule
      \textbf{Country} & \textbf{Languages} \\ \midrule
      United States & English, Spanish, Chinese, French, Undefined \\ 
      Russia & English, Russian, Thai\\ 
      China & English, Chinese, Thai, Spanish \\
      Antarctica & English, Thai, Undefined \\
      Kazakhstan & Russian, English \\
      Indonesia & Indonesian, English, French, Undefined \\
      Niger & English, Spanish \\
      Mali & English, Spanish, Tagalog, Hindi \\
      Algeria & English, French \\
    \bottomrule 
    \end{tabular}
    \caption{Languages posts are authored in for the countries where $<80\%$ of the bot population write in the dominant language, in descending order of frequency used}
    \label{tab:countries_languages_less_main}
\end{table}

\begin{table}[h]
    \centering
    \begin{tabular}{p{3cm}p{5cm}}
        \toprule
         \textbf{Region} & \textbf{Dominant Hashtags} \\ \midrule
       United States & COVID19Vaccine, AmericanRescuePlan, WearAMask, Biden, China \\ 
       Canada & cdnpoli, COVID19vaccine, VoteFordOut2022, paidsickdays, ltcjustice\\ 
       Russia & SputnikV, AIMS, AWANIpagi, NormaBaharu, education\\
       Africa & VaccineEquity, CR17BankStatements, HIV, Hennie, vaccines\\
       India & CoronaVaccine, NarendraModi, IndiaFightsCorona, AIMS, CovidFreeOdisha\\ 
       Indonesia & VaksinasiAnggotaPolri, polripresisi, AWANInews, HapusCOVID19, TempoMetro\\
       China & UnrestrictedBioweapon, vaccine, CCPVirus, ContemporaryGeneticWeapon, Sinovac\\
       Asia & LindungDiriLindungSemua, China, restrictions, COVID19Vaccine, Travel\\ 
        \bottomrule
    \end{tabular}
    \caption{Dominant Hashtags that the bots affiliated to each region are using}
    \label{tab:dominant_hashtags_main}
\end{table}

\section*{Methodology}
\subsection*{Data Collection}
For the main data of analysis, the coronavirus dataset, we manually selected a set of hashtags to collect tweets posted to the social media platform X. This list contained general terms related to the coronavirus vaccine, such as \#covidvaccine, \#coronavirusvaccine. We performed this data collection using the Twitter Streaming API V1 for the first five days of each month for a year through January to December 2021.

To identify bots across multiple languages, we built Multilingual BotBuster, a bot detection module that works across multiple languages. To identify the location affiliation of a user, we constructed a Geolocation Identifier module to discern the affiliated location of the social media users. This module referenced a gazetteer of locations and their coordinates using sources from Geonames and consolidated country coordinates. 

Finally, to determine the relationship between the cyber and physical space, we used auxiliary data such as the Gross Domestic Product (GDP) and the list of dominant languages of each country. This data were obtained from the World Bank and Wikipedia collections.

Details on the collection of all the data and the collection strategy are in the Appendix.

\subsection*{Multilingual Bot Detection}
Strategies to label social media users as bots or humans ranged from feature-based approach to machine learning approaches \cite{ng2022stabilizing,yang2022botometer}. Many of these bot detection algorithms were trained on posts written in the English language. However, our dataset reflected that only 47.4\% of the posts are written in English. This statistic was retrieved from the language code tagged to the tweet, indicating the language that X identifies the tweet to be in. With English-based bot detection algorithms, we were unable to discern whether the other agents were bots or humans. To analyze posts across a wide range of vernaculars, we needed a multilingual bot detection algorithm. 

We begin by creating a multilingual dataset. We translated three English-language datasets \cite{mazza2019rtbust,yang2019arming,cresci2019cashtag} that were manually annotated for bot/human users into three other languages: Chinese, Russian and Arabic. These languages were chosen because past literature suggests that bots from these countries are active in influencing online narratives \cite{jacobs2023tracking,bail2020assessing,simchon2022troll,albadi2019hateful}. 

Our multilingual bot detection algorithm was extended from the feature-based approach of the BotBuster detection algorithm \cite{ng2023botbuster}, and was therefore named \emph{``Multilingual BotBuster"}. We used three types of model architectures and tested the architectures on all the dataset language variations. The model architectures were: (1) language-specific model, (2) multilingual model, and (3) Large Language Models (LLM). For language-specific models, two to three variants of pretrained language models for each language were tested. Examples of these models were: google-bert/bert-base-uncased, google-bert/bert-base-chinese, blinoff/roberta-base-russian-v0, asafaya/bert-base-arabic. For multilingual model, four variants were tested. These variants were: TF-iDF vectorizer with random forest classifier, google-bert/bert-base-multilingual-uncased, FacebookAI/xlm-roberta-base and FacebookAI/xlm-mlm-enfr-1024. For the LLM architecture, two models were tested with six prompting schemes each. The models were declare-lab/flan-alpaca-gpt4-xl and openai-community/gpt2. We compared this set of models against four baseline models: BotBuster\cite{ng2023botbuster}, BotBuster For Everyone\cite{ng2024assembling}, Botometer\cite{yang2022botometer} and BotHunter\cite{beskow2018bot}.

From all the model variations, we selected the best performing model to be incorporated into the BotBuster algorithm. This model used google-bert/bert-base-multilingual-uncased\cite{DBLP:journals/corr/abs-1810-04805} as its tokenizer and pretrained model. The \emph{Multilingual BotBuster} algorithm performed with an accuracy of $82.79\pm4.23$\% across the languages tested (English, Chinese, Russian, Arabic).
Finally, we manually scrutinized a 1\% random sample of algorithmically inferred bot and human labels for a sanity check. 

The details of the Multilingual BotBuster are described further in the Appendix. 
With this multilingual bot detection algorithm, we had more analysis coverage of the actors, and therefore can better link the social actors to physical geography.

\subsection*{Geolocation Identifier}
The country of an agent reflects the affiliation of the agents. The cloak of anonymity of social media affords self-declaration of affiliation, and therefore identifying an agent's geolocation provides inference towards agent affiliation.

Current methods of geolocation typically use the Nomantim API or Large Language Models\cite{ng2023deflating,haupt2015conflict}. These methods can be slow because it relies on API calls, some of which have rate limits. To be able to identify locations at scale within our massive dataset, we built a gazetter-based \emph{Geolocation Identifier} script to identify geolocation coordinates from the metadata of an agent. This script first used the Stanford Named Entity Recognition parser \cite{finkel2005incorporating} to extract location words from the agent's self-written user description. Next, the script performed fuzzy matching using Levenshtein distance algorithm to retrieve the closest-matched location against a gazetteer. This gazetteer was built from a few open-sourced location databases\footnote{\url{https://public.opendatasoft.com/explore/dataset/geonames-all-cities-with-a-population-1000/table/?flg=en-us&disjunctive.cou_name_en&sort=name},\url{https://github.com/annexare/Countries}}. Our gazetter-based geolocation identifier can be run on a CPU, and is thus also accessible to be used without a GPU.

\paragraph*{Social Cyber Geography Analysis}
Social Cyber Geography relates the physical geography with social communities that arise in the cyber space. Our communities were differentiated by the bot agents' self-declared geographic location and their language used. Combined together, the patterns of geographic affiliation and linguistic targets reflects the communicative strategies of the bot-authored messages.

\emph{Bot Percentage by Country}
To understand the geographical distribution of bot affiliations, we identified the country of each tweet using the Geolocation Identifier script. We grouped the tweets by each identified country, and extracted the unique authors. Then, we calculated the ratio of the number of bots affiliated with each country against the total number of bots. This geographic heat map aggregated the bot percentage across the year's worth of data. We used the median as the averaging function, because across the bot percentage can have variations that disproportionately skewed the mean value across the year. The heatmap was presented in \autoref{fig:bot_perc_country}. 

\emph{Bot Percentage by Language}
To understand the target users of the bot agents, we identified the language of each tweet based on the language tagged in the tweet's metadata. This is an annotation provided by X. We grouped the tweets by language and extracted the unique authors. We calculated the bot percentage per language. \autoref{fig:bot_perc_language} presents the 30 languages with the highest bot percentage across the year's data.

\emph{Bot Percentage by Dominant Language of Country}
To harness the structure of the use of language by bots affiliated with each country, we calculated the percentage of bots affiliated with a country that had authored posts in the country's dominant language. We extracted the dominant languages of each country from Wikipedia. For each country, we compared the languages of the tweets affiliated to that country with the dominant languages. We then extracted the number of bot agents that authored tweets in the dominant language against the total number of bot agents affiliated with that country. The results are presented in a geographic heat map in \autoref{fig:bot_perc_lang_dominant}. For countries where less than 80\% of the bot population write in the country's dominant language, we examined the other languages that the posts were written in, and presented the results in \autoref{tab:countries_languages_less_main}.

\emph{Bot Percentage vs Country Indicators}
Finally, for the countries that bot data were present, we investigated the correlation between bot geographical distribution with two country related data. The first data was the economic activity, measured by the country's GDP. The second data was the population of the country. These country indicator data were extracted from the World Bank website. For each country, we performed a linear regression between the bot percentage and the country indicator data. This resulted in an $R^2=0.021$ between bot percentage and GDP and an $R^2=0.022$ between bot percentage and population. The low $R^2$ values indicated that there is no significant correlation between bot percentage and country indicators.
For a better examination of the narratives shared, we examined the dominant hashtags from the posts that are geolocated to each region, and presented the results in \autoref{tab:dominant_hashtags_main}.

\section*{Acknowledgments}
The research for this paper was supported by the following grants: Cognizant Center of Excellence Content Moderation Research Program, Office of Naval Research (Bothunter, N000141812108) and Scalable Technologies for Social Cybersecurity/ARMY (W911NF20D0002). The views and conclusions are those of the authors and should not be interpreted as representing the official policies, either expressed or implied.

\section*{Supplementary Material}
\subsection*{Multilingual BotBuster}
One of the key tasks in understanding bot activity across different geographical locations and languages is to identify bot accounts from tweets that are written in different languages.
Many of the current bot detection literature are based mostly on the English language. In this work, we extend the BotBuster detection algorithm \cite{ng2023botbuster} towards a Multilingual BotBuster algorithm, which aims to be able to analyze a variety of languages. 

In building the multilingual BotBuster, we first need to obtain data of several languages. We translated data from three manually annotated bot datasets from the OSOME repository: cresci-rtbust-2019 \cite{mazza2019rtbust}, botometer-feedback-2019 \cite{yang2019arming} and cresci-stock-2018 \cite{cresci2019cashtag}. We use Google Translate to translate the texts of the tweets from English to three other languages -- Chinese, Russian and Arabic. Following which, we performed bot classification using different types of pre-trained tokenizers and classifiers on the texts. These tokenizers and classifiers are pre-trained models from the HuggingFace repository. We used the default parameters for the classifiers, and used the same tokenizers to embed the tweet texts. In applying these algorithms, we used a 80:20 train:test stratified split to the dataset and ran a five-fold cross validation to determine algorithm accuracy. These runs are compared against baseline algorithms, which are algorithms that are commonly and commercially ran for social media bot detection.

For the baseline algorithms, we used: BotBuster\cite{ng2023botbuster}, BotBuster For Everyone\cite{ng2024assembling}, Botometer\cite{yang2022botometer} and BotHunter\cite{beskow2018bot}. These are algorithms that are commonly used in the bot detection literature that identify bots from X. 

In total, we tested three types of models: (1) language-specific models, (2) multilingual models and (3) Large Language Models. 

For language-specific models, we tested 2 to 3 classifier types of each language, except when not available. For the English language, we tested: distilbert/distilbert-base-uncased, FacebookAI/roberta-base, google-bert/bert-base-uncased. For the Chinese language, we tested google-bert/bert-base-chinese, hfl/chinese-roberta-wwm-ext. For the Russian language, we used blinoff/roberta-base-russian-v0. For the Arabic language, we used CAMeL-Lab/bert-base-arabic-camelbert-mix-pos-egy, asafaya/bert-base-arabic.

For the multilingual classifier, we tested 4 variations. The first is the TF-iDF vectorizer with a random forest classifier. The next three are taken from the HuggingFace model repository: google-bert/bert-base-multilingual-uncased, FacebookAI/xlm-roberta-base and FacebookAI/xlm-mlm-enfr-1024.

For Large Language Models, we tested a total of 6 prompts, each inserting the tweet text as the \{Sentence\}. These prompts were tested on two models: flan-alpaca-gpt4-xl and gpt2. The gpt2 model performed better than the flan-alpaca-gpt4-xl model. However, the outputs of the gpt2 model seemed unaffected by the prompt structure.

\begin{itemize}
    \item Prompt 1: ``````Do you think this sentence is written by a bot or a human? Output only the class `bot' or `human'. \{Sentence\} """
    \item Prompt 2: ``````Does this tweet look like it was written by a bot or a human? Output only the class `bot' or `human'. \{Sentence\} """
    \item Prompt 3: ``````Bots are automated actors on social media platforms. Does this tweet look like it was written by a bot or a human? Output only the class `bot' or `human'. \{Sentence\} """
    \item Prompt 4: ``````This sentence is in \{Language\} language. Do you think this sentence is written by a bot or a human? Output only the class `bot' or `human'. \{Sentence\} """
    \item Prompt 5: ``````This sentence is in \{Language\} language. Does this tweet look like it was written by a bot or a human? Output only the class `bot' or `human'. \{Sentence\} """
    \item Prompt 6: ``````This sentence is in \{Language\} language. Bots are automated actors on social media platforms. Does this tweet look like it was written by a bot or a human? Output only the class `bot' or `human'. \{Sentence\} """
\end{itemize}

Finally, we selected the best performing multilingual model (google-bert/bert-base-multilingual-uncased) to be used to identify bot and human accounts. This model is then ran on the collection of coronavirus tweets. The model returns a probability score between 0 and 1, which indicates the likelihood of the user being a bot. We segregate bot and human accounts using a threshold of 0.5, one that has been determined through previous longitudinal studies in bot detection literature \cite{ng2023botbuster}.
This Multilingual BotBuster model, in fact, performs better than most of the baseline algorithms for all the languages. Such an observation could be due to several factors: (1) the original algorithm was fine-tuned for the English language, and hence is unable to accurately detect bot accounts written in other languages; (2) the multilingual BERT-based model is more superior at understanding textual nuances as compared to the original algorithm. Such an improvement in the algorithm performance highlights the importance in continued improvement of bot detection algorithms as a research area. 

\begin{landscape}
\begin{table}
    \centering
    \begin{tabular}{p{5cm}cccccccccccc}
         ~ & \multicolumn{4}{l}{\textbf{cresci-rtbust-2019}} & \multicolumn{4}{l}{\textbf{botometer-feedback-2019}} & \multicolumn{4}{l}{\textbf{cresci-stock-2018}}  \\ \hline 
         ~ & English & Chinese & Russian & Arabic & English & Chinese & Russian & Arabic & English & Chinese & Russian & Arabic \\ \hline 
         \multicolumn{13}{l}{\underline{\textbf{Baseline Comparisons}}} \\ 
         BotBuster \cite{ng2023botbuster} & 67.20 & ~ & ~ & ~ & 54.20 & ~ & ~ & ~ & \textbf{79.35} & ~ & ~ & ~ \\
         BotBuster For Everyone \cite{ng2024assembling} & \textbf{70.65} & ~ & ~ & ~ & \textbf{83.08} & ~ & ~ & ~ & \textbf{79.35} & ~ & ~ & ~ \\ 
         Botometer \cite{yang2022botometer} & 60.10 & ~ & ~ & ~ & 53.68 & ~ & ~ & ~ & 38.12 & ~ & ~ & ~ \\ 
         BotHunter \cite{beskow2018bot} & 62.90 & ~ & ~ & ~ & 74.10 & ~ & ~ & ~ & 37.36 & ~ & ~ & ~ \\ \hline 
         \multicolumn{13}{l}{\underline{\textbf{Language-specific classifier}}} \\ 
         \multicolumn{13}{l}{\textbf{English}} \\ 
         distilbert/distilbert-base-uncased \cite{Sanh2019DistilBERTAD} & 90.12 & ~ & ~ & ~ & 88.49 & ~ & ~ & ~ & \textbf{91.09} & ~ & ~ & ~ \\ 
         FacebookAI/roberta-base \cite{DBLP:journals/corr/abs-1907-11692} & 83.55 & ~ & ~ & ~ & 85.71 & ~ & ~ & ~ & 78.40 & ~ & ~ & ~ \\
         google-bert/bert-base-uncased \cite{DBLP:journals/corr/abs-1810-04805} & \textbf{90.43} & ~ & ~ & ~ & \textbf{89.08} & ~ & ~ & ~ & 70.14 & ~ & ~ & ~ \\ 
         \multicolumn{13}{l}{\textbf{Chinese}} \\ 
         google-bert/bert-base-chinese \cite{huggingfaceGooglebertbertbasechineseHugging} & ~ & 86.11 & ~ & ~ & ~ & 71.93 & ~ & ~ & ~ & 78.68 & ~ & ~ \\
         hfl/chinese-roberta-wwm-ext \cite{chinese-bert-wwm} & ~ & \textbf{90.32} & ~ & ~ & ~ & \textbf{86.77} & ~ & ~ & ~ & 93.21 & ~ & ~ \\
         \multicolumn{13}{l}{\textbf{Russian}} \\ 
         blinoff/roberta-base-russian-v0 & ~ & ~ & \textbf{85.56} & ~ & ~ & ~ & \textbf{80.78} & ~ & ~ & ~ & \textbf{50.20} & ~ \\
         \multicolumn{13}{l}{\textbf{Arabic}} \\ 
         CAMeL-Lab/bert-base-arabic-camelbert-mix-pos-egy \cite{inoue-etal-2021-interplay} & ~ & ~ & ~ & 88.98 & ~ & ~ & ~ & 87.39 & ~ & ~ & ~ & 77.98 \\ 
         asafaya/bert-base-arabic \cite{safaya-etal-2020-kuisail} & ~ & ~ & ~ & \textbf{90.62} & ~ & ~ & ~ & \textbf{89.40} & ~ & ~ & ~ & \textbf{84.68} \\ \hline
         \multicolumn{13}{l}{\underline{\textbf{Multilingual classifier}}} \\ 
         TF-iDF vectorizer + random forest classifier & 80.87 & 79.06 & 80.12 & 79.02 & 66.19 & 67.42 & 67.72 & 67.55 & 78.53 & 70.65 & 77.99 & 74.52 \\
         google-bert/bert-base-multilingual-uncased \cite{DBLP:journals/corr/abs-1810-04805} & \textbf{87.74} & \textbf{91.04} & \textbf{82.59} & \textbf{82.83} & \textbf{85.63} & \textbf{81.95} & \textbf{83.99} & \textbf{84.25} & 79.13 & 79.06 & 80.01 & \textbf{75.26} \\ 
         FacebookAI/xlm-roberta-base \cite{DBLP:journals/corr/abs-1911-02116} & 78.18 & 83.33 & 79.10 & 76.19 & 78.97 & 78.97 & 76.03 & 79.65 & 65.94 & 57.93 & 53.97 & 73.71 \\
         FacebookAI/xlm-mlm-enfr-1024 \cite{lample2019cross} & 81.63 & 83.40 & 77.09 & 75.24 & 74.78 & 69.53 & 67.38 & 66.66 & \textbf{79.53} & 79.53 & 78.04 & 66.53 \\ \hline 
         \multicolumn{13}{l}{\underline{\textbf{Large Language Models}}} \\ 
         \multicolumn{13}{l}{\textbf{declare-lab/flan-alpaca-gpt4-xl \cite{bhardwaj2023red}}} \\
         prompt 1 & 31.69 & 12.17 & 10.65 & 2.12 & 40.96 & 14.28 & 20.87 & 4.36 & 27.20 & 16.99 & 16.74 & 4.79 \\ 
         prompt 2 & 43.03 & 20.57 & 22.04 & 8.63 & 52.77 & 21.21 & 27.90 & 7.33 & 35.64 & 22.64 & 22.24 & 18.66 \\ 
         prompt 3 & 26.28 & 16.94 & 10.97 & 33.43 & 35.64 & 19.12 & 16.87 & 4.13 & 22.84 & 20.45 & 13.54 & 4.96 \\ 
         prompt 4 & 30.47 & 10.42 & 2.80 & 5.35 & 45.48 & 11.99 & 7.35 & 5.86 & 34.27 & 17.41 & 6.58 & 10.47 \\
         prompt 5 & 39.28 & 14.14 & 4.55 & 15.23 & 57.17 & 15.29 & 6.90 & 11.47 & 43.61 & 18.80 & 6.73 & 13.23 \\ 
         prompt 6 & 19.73 & 20.45 & 1.61 & 5.65 & 28.98 & 25.03 & 5.92 & 6.85 & 25.07 & 23.91 & 6.94 & 8.35 \\
         \multicolumn{13}{l}{\textbf{openai-community/gpt2 \cite{radford2019language}}} \\ 
         prompt 1 & 66.22 & 67.55 & 65.32 & 63.95 & 87.32 & 84.06 & 88.26 & 89.50 & 67.99 & 70.12 & 64.07 & 69.71 \\ 
         prompt 2 & 66.22 & 67.55 & 65.32 & 63.95 & 87.32 & 84.06 & 88.26 & 89.50 & 67.99 & 70.12 & 64.07 & 69.71 \\ 
         prompt 3 & 66.22 & 67.55 & 65.32 & 63.95 & 87.32 & 84.06 & 88.26 & 89.50 & 67.99 & 70.12 & 64.07 & 69.71 \\ 
         prompt 4 & 66.22 & 67.55 & 65.32 & 63.95 & 87.32 & 84.06 & 88.26 & 89.50 & 67.99 & 70.12 & 64.07 & 69.71 \\ 
         prompt 5 & 66.22 & 67.55 & 65.32 & 63.95 & 87.32 & 84.06 & 88.26 & 89.50 & 67.99 & 70.12 & 64.07 & 69.71 \\ 
         prompt 6 & 66.22 & 67.55 & 65.32 & 63.95 & 87.32 & 84.06 & 88.26 & 89.50 & 67.99 & 70.12 & 64.07 & 69.71 \\ \hline
    \end{tabular}
    \caption{F1 scores for multilingual classifiers}
    \label{tab:F1 Scores}
\end{table}
\end{landscape}

\subsection*{Geographical Location Identification}
To identify the geographical location from the tweets, we constructed a Geolocation Identifier Script. This script examines the metadata from each tweet to extract the location string. 
We chose to write our own script to provide offline geolocator capabilities of large amounts of data. \autoref{tab:geo_comparison} shows a few methods that we had compared before settling on the current method. 

The first method uses an LLM to extract locations within a description, prompting the LLM with: ``````Return the locations (country, city, place) that are stated within this sentence: \{sentence\}""". We used the open source declare-lab/flan-alpaca-large model (\url{https://huggingface.co/declare-lab/flan-alpaca-large}) to evaluate this prompt. The extracted locations is then sent to the Nomantim API to retrieve the location coordinates. While this method returns a good proportion of locations (60\% of users with description returned a location), it was rather slow, for it relied on the LLM processing and the Nomantim API calls. 

The second method is a Nomantim and OpenStreetMap combination. In this method, the entire user description string is sent to OpenStreetMap to retrieve a list of possible locations. Then, the list of locations is sent to Nomantim API to retrieve the location coordinates (latitude, longitude).

\begin{table*}
    \centering
    \begin{tabular}{p{3cm}p{4cm}p{4cm}p{4cm}}
         & \textbf{LLM + Nomantim} & \textbf{Nomantim + OpenStreetMap} & \textbf{Gazetteer comparison} \\ \hline 
         \textbf{Description} & Use an LLM to extract out locations within the description, and retrieving location coordinates from Nomantim API & Send the entire user description string to OpenStreetMap to retrieve possible locations, and retrieving location coordinates from Nomantim API  & Fuzzy comparison of entities extracted from user description with a consolidated gazetteer \\ 
         \textbf{Speed} & Slow & Slowest & Fastest \\
         \textbf{\% locations returned from users with description} & 60 & 56 & 62 \\ 
         %Past usages &  & \cite{ng2023deflating} & NA \\ 
         \textbf{Limitations} & Relies on LLM calls, which may require payment; Relies on Nomantim rate limit & Relies on Nomantim rate limits & Requires extensive gazetteer \\
         \hline 
    \end{tabular}
    \caption{Comparison of Geolocation Methods to identify location coordinates from user description}
    \label{tab:geo_comparison}
\end{table*}

The third method is a gazetter comparison method which we used. Our Geolocation Identifier Script stems from a gazetteer of locations consolidated from open-sourced information downloaded from the Internet. Namely, it consists of: (1) cities1000 from GeoNames, which consists of all cities that have a population of more than 1000 people. This data is extracted from the url \url{https://public.opendatasoft.com/explore/dataset/geonames-all-cities-with-a-population-1000/table/?flg=en-us&disjunctive.cou_name_en&sort=name}. (2) latitude and longitude data of countries downloaded from a Github repository that consolidated location information (\url{https://github.com/annexare/Countries}). This script takes in the user description in the form of a string and returns the country, the latitude and the longitude of possible locations.

To begin, we use the Stanford Named Entity Recognition parser to extract out location words within the agent's description \cite{finkel2005incorporating}. Next, we use the Python fuzzywuzzy library (\url{https://github.com/seatgeek/fuzzywuzzy}) to perform fuzzy matching using Levenshtein distance of the location names against the gazetteer. The names are matched in the following order: (1) city, country pair, (2) country, (3) city. If a the gazetteer returns a match of $>$0.80, that location is returned along with the latitude/ longitude coordinates.

This gazetter-based method is both the fastest running algorithm and returns the largest percentage of location coordinates. As it relies on pre-consolidated gazetter, this method is able to work in an offline setting unlike the methods with the API calls, and does not require extensive compute power unlike the LLM method. 

\section*{Investigations around Geographical Location of Bot Agents}
Identifying geographical locations from social media posts such as tweets provides insights to the interactions within the digital space. We term this ``social cyber geography", where the space in the cyber realm, which consists of information technology such as computers, social media and virtual reality, is produced through social relations. These social relations, be it a tweet or a retweet, reflect the unity of both cyber and physical geography, thereby reflecting both virtual and physical communities. This concept had been used in digital diplomacy studies \cite{ng2023deflating,jacobs2023tracking,pohan2016digital}, to reflect the harmony and the tensions between countries. 

\subsection*{Bot Percentage by Country}
We identified the country each tweet was affiliated with using the method detailed in the Geographical Location Identification portion. We then grouped the tweets per country, and calculated the bot percentage for each country. Then, we plotted a geographic heat map to reflect the percentage of bot users in each country present in the data. We plotted one chart per month, and an aggregated chart where we use the median as an averaging function. There were some countries where we did not retrieve any bot agents. We compared our data against the GeoCOV19 dataset for further validation \cite{qazi2020geocov19}. The GeoCOV19 dataset does not differentiate bot and human agents, but instead presents the number of tweets for these countries. We found the countries where there were no bot agents reasonable, because the GeoCOV19 dataset showed 0-100 tweets by all agents for the same regions. For these countries, the usage of X were rather small ($\sim$500k) too. For these countries, the usage of the platform X to discuss coronavirus discourse were rather small, thus might not be worth it for bots to target those regions.

\subsection*{Bot Percentage by Language}
We identified the language of each tweet based on the language tagged in the tweet's metadata. This language annotation is determined by X's internal machine learning algorithms.
We then grouped the tweets by language and calculated the bot percentage per language. Then we plotted a bar chart reflecting the languages that have the most highest average bot proportion, where we used the mean as the averaging function. We plotted one chart per month, and an aggregated chart.

\subsection*{Bot Percentage by Country's Dominant Language}
We retrieved each country's dominant language from the Wikipedia page \url{https://en.wikipedia.org/wiki/List_of_official_languages_by_country_and_territory}. For the set of bots that are affiliated with each country, we identified how many of them had tweets written in the same language(s) as the country's dominant language. We then plotted a geographic heat map to reflect the prominence of bots that post in each country's dominant language. We plotted one chart per month, and an aggregated chart where we use the mean as the averaging function.

\subsection*{Bot Percentage vs GDP and Population}
We retrieved the GDP (Gross Domestic Product) and Population per country from the World Bank. We then ran two linear regressions: the first of the percentage of bots affiliated with each country against the GDP, and the second of the percentage of bots affiliated with each country against the population of the country. We find no significant correlation between the bot percentage and the GDP or population of the country. The $R^2$ value for bot percentage and GDP is 0.021. The $R^2$ value of bot percentage and population is 0.022.

\begin{figure*}[tbhp]
\centering
\includegraphics[width=\linewidth]{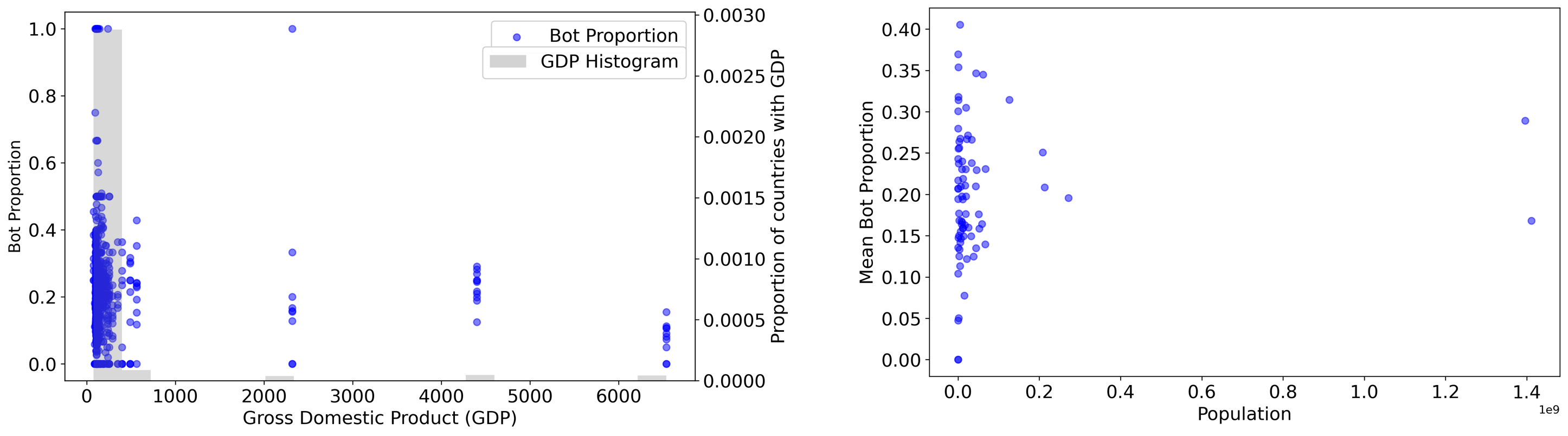}
\caption{Scatter Map of Mean Bot Proportion vs. (a) GDP and (b) Population of country}
\label{fig:bot_perc_gdp}
\end{figure*}

\section*{Results aggregated by month}
\begin{figure*}[tbhp]
\centering
\includegraphics[width=\linewidth]{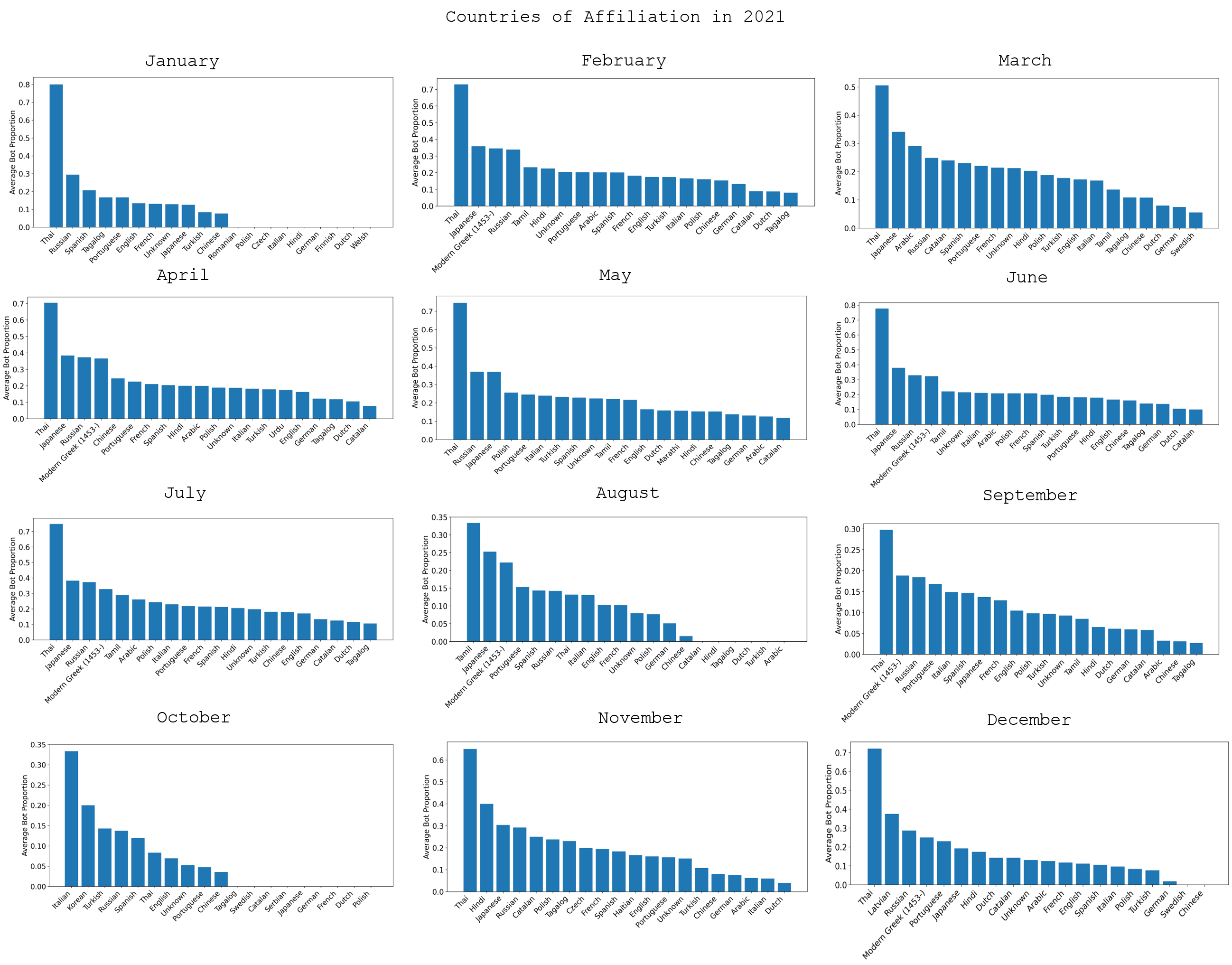}
\caption{Bot proportion against commonly used languages by month}
\label{fig:bot_perc_language_monthly}
\end{figure*}

\begin{figure*}[tbhp]
\centering
\includegraphics[width=\linewidth]{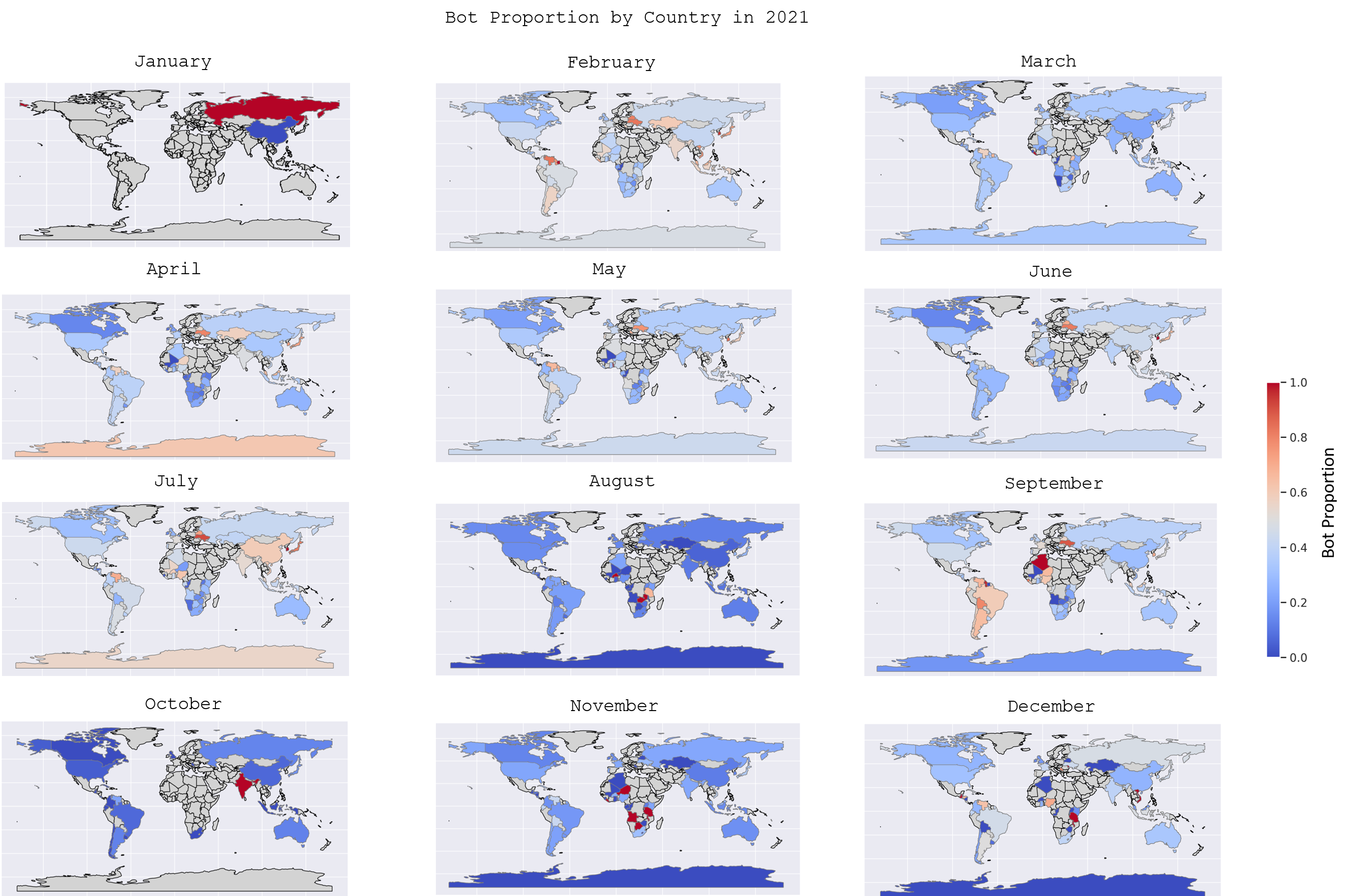}
\caption{Geographical heatmap of bot proportion affiliated to each country by month}
\label{fig:bot_perc_country_monthly}
\end{figure*}

\begin{figure*}[tbhp]
\centering
\includegraphics[width=\linewidth]{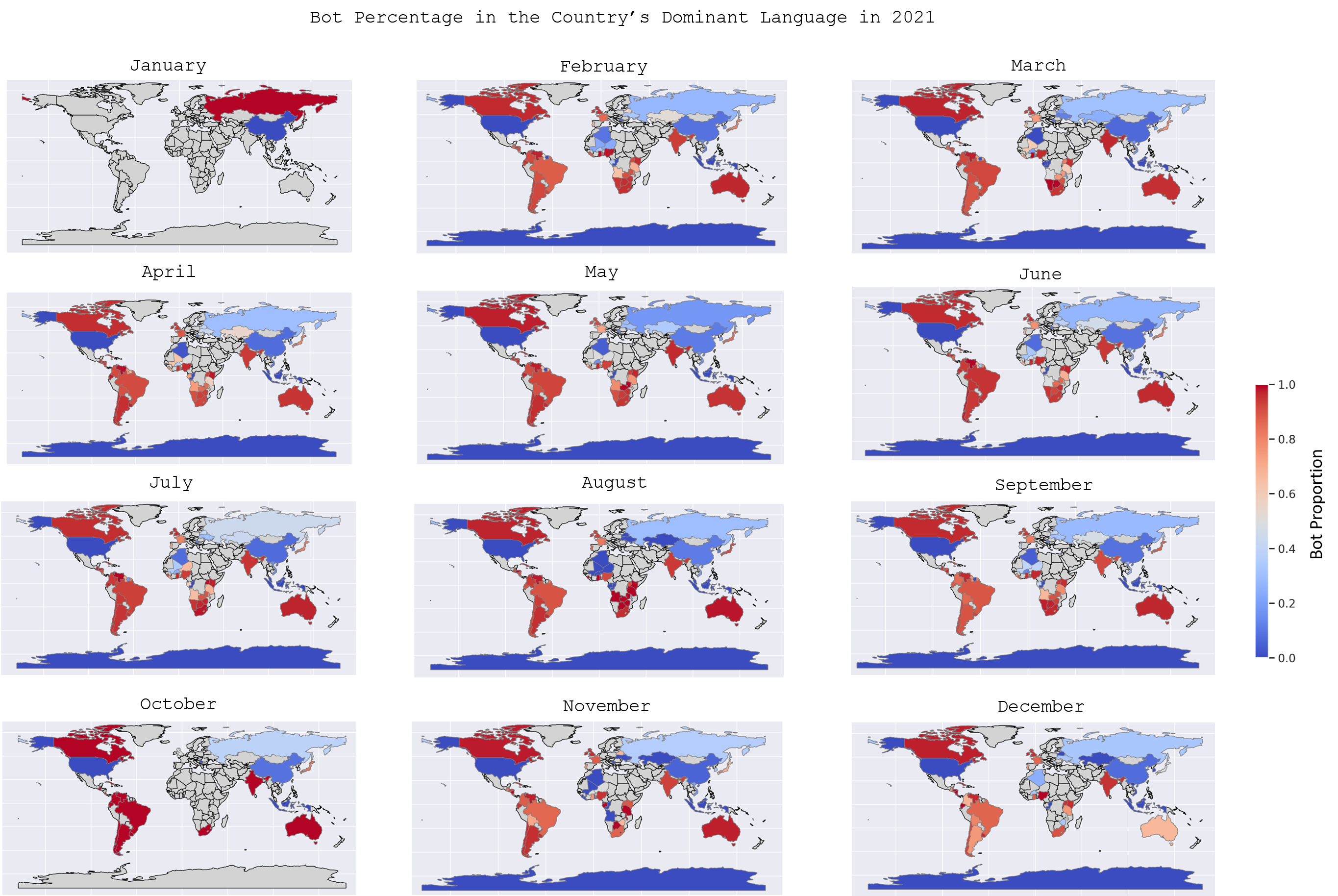}
\caption{Geographical heatmap of bot proportion authoring posts in the country's dominant language by month}
\label{fig:bot_perc_dominant_language_monthly}
\end{figure*}

\autoref{fig:bot_perc_language_monthly} presents the bot proportion against the most commonly used languages in the dataset per month. \autoref{fig:bot_perc_country_monthly} presents the geographical heatmap of bot percentage affiliated to each country by month. 
\autoref{fig:bot_perc_dominant_language_monthly} presents the geographical heatmap of bot percentage that is affiliated to each country and posting in the country's dominant language by month.

\section*{Comparison of Bot Percentage per country against GeoCOV19 dataset}
We compared our distribution of bot agents with the GeoCOV19 dataset \cite{qazi2020geocov19} for the countries that our dataset shows to have no bot-authored tweets. The GeoCOV19 dataset geolocated multilingual tweets over a period of 90 days. The dataset does not differentiate bot and human agents. \autoref{tab:comparison_with_geocov19} presented the comparison for these countries. Our results were in line with the GeoCOV19 dataset, because where the dataset showed 0-100 or 100-1000 tweets by all agents (bot and humans), our dataset showed no bot-authored tweets. This was a reasonable finding because the total number of tweets are already small, which means the bot subset might be smaller, or even none. For these countries, the usage of the platform X to discuss coronavirus discourse were rather small, thus might not be worth it for bots to target those regions.

\begin{table*}
    \centering
    \begin{tabular}{p{3cm}p{4cm}p{4cm}p{4cm}}
        \toprule
        \textbf{Country} & \textbf{\% of bots from Multilingual BotBuster}  & \textbf{Number of tweets from GeoCOV19} & \textbf{Estimated users of X} \\ \midrule 
        Greenland & 0 & 0-100 & 7600 \\ 
        Mexico & 0 & 100-1000 & 12.2 million \\ 
        Central Africa & 0 & 0-100 & 550,000 \\
        North Africa & 0 & 0-100 & 8 million \\ 
        Mongolia & 0 & 0-100 & 151,000 \\
        Thailand & 0 & 100-1000 & 14.2 million \\
        Myanmmar & 0 & 0-100 & 1.45 million \\
        Norway & 0 & 100-1000 & 486,000 \\
        Sweeden & 0 & 100-1000 & 273,528 \\
        \bottomrule
    \end{tabular}
    \caption{Comparison of number of bot users from our dataset against number of tweets from GeoCOV19 dataset, for countries where our dataset shows no bot-authored tweets. The estimated number of users of X were taken from the 2024 social media platform survey by Statista or DataReportal}
    \label{tab:comparison_with_geocov19}
\end{table*}

\section*{Languages for countries where $<80\%$ of bot proportion write in the dominant language}
For most countries, at least 80\% of the bots authored posts in the country's dominant language. \autoref{tab:countries_languages_less} shows a table which investigates the languages that the posts are authored in for the countries that have less than $80\%$ of the posts authored in the dominant language. This table is ordered in descending order of frequency the language is used.

\begin{table}
    \centering
    \begin{tabular}{p{3cm}p{5cm}}
    \toprule
      \textbf{Country} & \textbf{Languages} \\ \midrule
      United States & English, Spanish, Chinese, French, Undefined \\ 
      Russia & English, Russian, Thai\\ 
      China & English, Chinese, Thai, Spanish \\
      Antarctica & English, Thai, Undefined \\
      Kazakhstan & Russian, English \\
      Indonesia & Indonesian, English, French, Undefined \\
      Niger & English, Spanish \\
      Mali & English, Spanish, Tagalog, Hindi \\
      Algeria & English, French \\
    \bottomrule 
    \end{tabular}
    \caption{Languages posts are authored in for the countries where $<80\%$ of the bot population write in the dominant language, in descending order of frequency used}
    \label{tab:countries_languages_less}
\end{table}

\section*{Dominant Hashtags of Each Region}
We divided the world into a few pertinent regions based on the patterns of bot distribution, and understand the narratives through the hashtags used. These regions are: United States, Canada, Russia, China, Africa, India, Indonesia, and Asia.
For an understanding of the conversations in each region, we parsed the dominant hashtags contained within the posts authored by the bots that are affiliated to each region. We manually merged hashtags that are similar together, such as ``COVID19Vaccine" and ``COVID-19Vaccine", or the same phrase in different languages. We ignored the common hashtags ``COVID19" and ``Coronavirus" and their variations. \autoref{tab:dominant_hashtags} displays the top 5 hashtags for each region.

\begin{table}
    \centering
    \begin{tabular}{p{3cm}p{5cm}}
        \toprule
         \textbf{Region} & \textbf{Dominant Hashtags} \\ \midrule
       United States & COVID19Vaccine, AmericanRescuePlan, WearAMask, Biden, China \\ 
       Canada & cdnpoli, COVID19vaccine, VoteFordOut2022, paidsickdays, ltcjustice\\ 
       Russia & SputnikV, AIMS, AWANIpagi, NormaBaharu, education\\
       Africa & VaccineEquity, CR17BankStatements, HIV, Hennie, Vaccines\\
       India & CoronaVaccine, NarendraModi, IndiaFightsCorona, AIMS, CovidFreeOdisha\\ 
       Indonesia & VaksinasiAnggotaPolri, polripresisi, AWANInews, HapusCOVID19, TempoMetro\\
       China & UnrestrictedBioweapon, vaccine, CCPVirus, ContemporaryGeneticWeapon, Sinovac\\
       Asia & LindungDiriLindungSemua, China, restrictions, COVID19Vaccine, Travel\\ 
        \bottomrule
    \end{tabular}
    \caption{Dominant Hashtags of Each Region}
    \label{tab:dominant_hashtags}
\end{table}

\section*{Data Collection}
We collected our main set of data from X during the year of 2021. Specifically, we collected data from the first 5 days of each month. We used the Twitter Streaming API V1. These dataset is called ``Coronavirus dataset". In total we collected data from 31,815,191 unique users, spanning about 100 million tweets.

\subsection*{Hashtags for Collection}

The hashtags used for the Coronavirus dataset collection are:

VaccinesWork, Sharethevaccine, ProtectVaccineProgress, getvaccine, WaitforVaccine, FreeTheVaccine, vaccinesaresafe, vaccineconfidence, igotvaccinated, coronavaccineforall, CoronavirusVaccineAppointment, justtakethefuckingvaccine, VaccinesWithoutBorders, Vaccines4All, vaccinesaves, Vaccine4ALL, BreakthroughVaccine, waitingformyvaccine, GetTheVaccine, GetYourVaccine, takethevaccine, vaccinesafe, Vaccineswillwork, Iwilltakethevaccine, vaccinefreedom, VaccineToAll, SafeAndEffectiveCovid19Vaccine, VACCINESARESAFE, safevaccines, nosleeptilvaccine, NoOnsiteSchoolsUntilVACCINES, WhereIsMyVaccine, vaccineselfie, vaccineready, vaccinesareamazing, vaccinee, VirusToVaccine2020, HoHoHopeVaccineArrivesSoon, Vaccined, VaccinesWork, VaccinesSaveLives, Vaccine4All, effectivevaccine, VaccineForSA, VaccinesSavesLives, SafeVaccines, accesstovaccines, VaccineHope, VaccinesWithoutBorders, wherearethevaccines, VaccinesforAll, getthevaccine, giveusthevaccine, VaccineWorks, vaccinesavelifes, GetAVaccine, vaccinesafelife, safevaccine, HaveTheVaccine, WhyIGotMyVaccine, CovidVaccineforall, vaccinesavestheworld, ImGettingTheVaccine, FreeVaccines, VaccinesWorkforAll, GiveMeMyVaccineNow, Affordablevaccine4all, getthatvaccine, justgivemethevaccine, SayYestotheVaccine, TakeYourVaccine, provaccine, YesToVaccine, vaccinesave, covid19\_vaccine\_4all, vaccineissafe, PleaseGetTheVaccine, VaccinesForEveryone, TrustTheVaccine, vaccinessavelives, VaccinesWork, Vaccine4All, TakeTheVaccine, FirstDoseOfVaccine, We4Vaccine, vaccinesaveslives, VACCINEFORWELLNESS, TheVaccineIsSafe, IWillTakeTheVaccine, firstdosevaccine, IWillTakeVaccine, SafeVaccine, SayYesToVaccine, VaccinesWorkForAll, VaccineIsSafe, TrustTheVaccine, getyourvaccine, SafeVaccines, VaccineSavesLives, vaccinesaresafe, YesToCovid19Vaccine

NoVaccines, NoVaccinesForMe, VaccineYourAss, NoToCoronavirusVaccines, SayNoToVaccine, NoMandatoryVaccine, VaccinesKill, VaccinesKills, stopvaccine, fkyourvaccines, antivaccine, AntiVaccine, NoVaccine, StopCovidVaccine, WeDoNotConsentCVVaccine, VaccinesKill, VaccinesHarm, SayNoToVaccines, ForcedVaccines, VaccineIsPoison, ResistVaccines, Noneedvaccines, FakeCoronaVaccine, VaccinesAreBioweapons, Iwontgetthevaccine, VaccineFromHell, vaccinepoison, StopAllVaccines, Vaccinetakedown, vaccinesDAMANGEimmunity, vaccinesRnotNATURAL, RejectTheVaccines, BewareVaccines, FuckVaccines, HellNoVaccine, NoVaccinesEver, WhoNeedsVaccine, justsaynotoforcedvaccines, StickTheVaccineUpYourArse, nottakingavaccine, vaccinebad, WeaponizedDeadlyVaccines, JustSayNOToTheVaccines, DoNotTakeTheVaccine, FuckVaccinePoison, NOVaccine4Me, VaccinesNotSafe, VaccineBioWeapon, wedontwantvaccine, vaccinenotforme, DontTakeCovidVaccine, NotTakingCovidVaccine, DangerousVaccine, vaccinedoesnotwork, VaccineIsntSafe, No2Vaccine, OpposeTheVaccine, YouCanHaveMyVaccine, ShoveThatVaccineUpYourAsshole, MoThankYouCovidVaccine, ToHellWithCovidVaccine, CovidVaccinePoison, SCREWTHEVACCINE, murdervaccine, VaccineIsUseless, KilloronaVaccine, DodgyVaccine, DoNotTakeAnyVaccine, JeNeMeVaccineraiPas, NoVaccineForMe, stopthevaccine, novaccine, BoycottIndianVaccine, vaccinehesitancy, To\_Vaccine\_Is\_My\_Choice, vaccinefailure, donttakethevaccine, FakeVaccine, StolenVaccines, NoVaccine4Me, TrudeauVaccineContractsLie, vaccinesKill, VaccinesArePoison, VaccinesAreNotCures, NotAVaccine, jesusisvaccine, HALTtheVaccines, NoVaccines, dontgetthevaccine, VaccinesHarm, TeamNoVaccine, TheVaccineIsTheVirus, fakeVaccines, killervaccine, lethalvaccine, CoronaVaccineFail, vaccineBioweapon, saynotovaccines, Fuckvaccines, fuckyourvaccines, vaccinedeath, JustSayMoToVaccines, NoCOVIDVAccineMandate, NoMandatoryVaccines, GoHomeLeaveOurVaccinesAlone, antivaccine, anti\_vaccine, stopcovidvaccine, CovidVaccineHesitancy, VaccinesCanKill, Antivaccines, notovaccine, DontGetVaccine, VaccineDontWork, VaccineKills, StopVaccine, ForcedVaccine, FakeCovidVaccine, VaccinesAreBad, Falsevaccine, PfizerVaccineKillingPeople, NoCovidVaccineForMe, VaccineShaming, TakeTheVaccineAndShoveItUpYourAss, FakeVaccinesWillNotSaveYou, WorldSaysNoVaccine, NOCovidVaccine, NOCovid19Vaccine, Jesusovervaccines, IAmNOTVAccineBait, NotAVaccineAMedicalExperiment, VaccineMortality, NoVaccine, NoVaccineForMe, NOVACCINE4ME, VaccineDeaths, VaccineDeath, VaccineInjury, AntiVaccine, vaccinesideeffect, NoToCoronaVirusVaccines, AvoidCovidVaccine, NoVaccines, coronavirusvaccinescam, ForcedVaccines, destroyvaccines, VaccineHesitancy, noneed4vaccine, notocovidvaccine, vaccinesharm, vaccinedamage, vaccinesareevil, SayNoToVaccines, killervaccine, no2vaccine, vaccinekills, vaccineskill, VaccinesKillingpeople, JustSayNotoVaccines, vaccinedanger, donttrustthecovid19vaccine, RejectWeaponizedVaccines, StopVaccine, FakeVaccine, vaccinechemicalweapon, DeathToVaccines, VaccineScam, NoVaccinesNeeded, NoVaccinesForMe, vaccinehesitant, PoisonVaccine, DeathVaccine, Cancervaccine, VaccineFail, vaccineRESISTANT, VaccineNonsense, UnsafeVaccines, NoVaccinesNecessary, AbnominableVACCINE, Vaccinefuckup, novaccinerequired, vaccinedrivemutations, VaccineFromHell, GodIsMyVaccine, AntiCovidVaccine, NeitherDoTheseCOVIDVaccines, notovaccine, Notmyvaccine, CovidVaccineIsPoison, VaccineDisaster, NovavaxVaccine, TheCovidVaccineKills, vaccinekills, StopTheVaccines, fuckyourvaccine, VaccineIssues, vaccinesDONTwork, vaccinebad, VaccineNotTheAnswer, dontgetthecovidvaccine, MurderbyVaccine, vaccineforwhat, norealvaccine, NoVaccine, VaccineScam, VaccineDeaths, NotAVaccine, VaccineKills, NOVACCINE4ME, vaccineisdeath, vaccineispoison, NoVaccineForMe, To\_Vaccine\_Is\_My\_Choice, vaccinesharm, NoVaccinePassports, NoToVaccinePassports, VaccinesAreDangerous, VaccinesKill, VaccinesAreBad, saynotovaccines, FuckTheVaccine, StillAintGettingTheVaccineThough, ChineseVaccineBioWeapon, antivaccine, JustSayNotoVaccines, Notomandatoryvaccines, fakevaccines, vaccinedeath , vaccineFail, StopVaccinePassports, CovidvaccineFail, COVID19VaccinesBioWeaponsOfMassDestruction, NoMandatoryVaccines, DoNotGetTheVaccine

\section*{Data Availability}

\subsection*{Terms of Twitter Data Access}
To allow reproduction and extension of our findings, we make the data available upon request, by emailing the corresponding author. The data shared will be in accordance to the Twitter Developer Agreement and Policy. Further sharing is not permitted, per the Twitter legal policies.
Researchers will be required to sign up for a Twitter Developer Account should they want to collect a full set of the data.

\subsection*{Third party data}
The data used for constructing the geolocator can be obtained from the following sites: (1) the geonames and geolocation (latitude, longitude) of cities in the world with a population greater than 1000 people \url{https://public.opendatasoft.com/explore/dataset/geonames-all-cities-with-a-population-1000/table/?flg=en-us&disjunctive.cou_name_en&sort=name} and (2) the geolocation of countries and their abbreviations \url{https://github.com/annexare/Countries}.

The 2020 GDP and Population data was obtained from the World Bank Data Bank GDP deflator at: \url{https://databank.worldbank.org/metadataglossary/world-development-indicators/series/NY.GDP.DEFL.ZS}.

The dominant language of each country was obtained from Wikipedia at \url{https://en.wikipedia.org/wiki/List_of_official_languages_by_country_and_territory}.

\bibliography{biblography}

\end{document}